\documentclass[11pt,a4paper]{article}
\usepackage{amsmath,amssymb}
\usepackage{epsfig,graphicx}
\usepackage[section]{placeins}
\begin{document}
\title{\vskip -70pt
\begin{flushright}
{\normalsize DAMTP-2008-79}
\end{flushright}
\vskip 50pt
\bf Light Nuclei as Quantized Skyrmions:\\Energy Spectra and Form Factors}
\author{N.~S.~Manton\footnote{{\bf e-mail}: N.S.Manton@damtp.cam.ac.uk}
~\,
and
~
S.~W.~Wood\footnote{{\bf e-mail}: S.W.Wood@damtp.cam.ac.uk}
\\
\small{\em DAMTP, University of Cambridge,} \\
\small{\em Wilberforce Road, Cambridge CB3 0WA, United Kingdom}
}
\date{}
\maketitle

\begin{abstract}
We review the $SU(2)$ Skyrme model and describe its topological soliton
solutions, which are called Skyrmions. Skyrmions provide a model of nuclei in which the conserved 
topological charge is identified with the baryon number of a nucleus.
The semiclassical quantum theory of Skyrmions, in which they are treated as rigid
bodies spinning in space and isospace, is described. 
We derive the energy spectra corresponding to various light nuclei, 
and predict a few new states.
We also calculate
the electromagnetic form factors describing the structure of the $\alpha$-particle and
lithium-6. Our recent reparametrization of the model gives results
that are in reasonable quantitative agreement 
with experiment. 

\end{abstract}
\section{Introduction}
The Skyrme model, a nonlinear classical field theory of 
pions, was introduced in the early 1960s as a tentative description of the strongly interacting 
elementary particles \cite{skyrme}. The model provides a low energy effective theory of
quantum chromodynamics (becoming exact as the number of quark colours becomes large) \cite{wittenglobal,wittencurrent}.
QCD has an important
broken symmetry: the approximate chiral $SU(2) \times SU(2)$ symmetry of strong
interactions. By considering the conserved vector and axial-vector currents of QCD, it can be deduced that if
this symmetry is exact and unbroken, parity doubling would be seen in the
hadron spectrum. However, no such phenomenon is observed. So
chiral symmetry must be spontaneously broken to its
isospin $SU(2)$ subgroup. 
A spontaneously broken approximate chiral symmetry 
entails the existence of approximately massless Goldstone bosons. The three pion
particles $\pi^+$, $\pi^0$ and $\pi^-$ behave as these approximate Goldstone bosons \cite{weinberg}.
The Skyrme model captures this broken chiral symmetry.

Interestingly, the Skyrme model admits topological soliton
solutions, \emph{Skyrmions}, with an integer-valued conserved topological charge.
Skyrmions provide a model of atomic nuclei in which one interprets a quantized charge $B$ Skyrmion
as a nucleus with baryon number $B$.
We describe a semiclassical quantum theory of
Skyrmions and our recent progress in describing light nuclei within this framework.

\section{The Skyrme Model}
The Skyrme model is a nonlinear theory
of pions defined in terms of four fields: $\sigma$, $\pi_1$, $\pi_2$ and 
$\pi_3$, subject to the constraint $\sigma^2 + \pi_1^2 + \pi_2^2 + \pi_3^2=1$ \cite{manton}. 
The Skyrme field is an $SU(2)$ matrix defined as
\begin{equation}
U = \sigma 1_2 + i{\boldsymbol{\pi}}\cdot {\boldsymbol{\tau}}=\left(
\begin{array}{cc}
\sigma+i\pi_3 & i\pi_1+\pi_2 \\
i\pi_1-\pi_2 & \sigma-i\pi_3
\end{array}
\right)\,,
\end{equation}
where $\boldsymbol{\tau}$ denotes the triplet of Pauli matrices.

The Lagrangian density is given by
\begin{equation}
\mathcal{L} = \frac{F_\pi^2}{16}\,\hbox{Tr}\,\partial_\mu U
\partial^\mu U^{\dag} + \frac{1}{32e^2}\,\hbox{Tr}\,[\partial_\mu U U^{\dag},
\partial_\nu U U^{\dag}][\partial^\mu U U^{\dag},
\partial^\nu U U^{\dag}] + \frac{1}{8} m_\pi ^2 F_\pi^2\,\hbox{Tr}\,(U-1_2) \,,
\end{equation}
where $F_\pi$ is the pion 
decay constant, $e$ is a dimensionless parameter and $m_\pi$ is the pion mass. 
One imposes the boundary condition $U({\mathbf{x}}) \rightarrow 1_2$ as $|{\mathbf{x}}|\rightarrow
\infty$. The vacuum, the unique field of minimal energy, is then $U({\mathbf{x}}) = 1_2$
for all $\mathbf{x}$. In the absence of the term involving the pion mass, the Lagrangian would be symmetric under the
$SU(2) \times SU(2)$ chiral symmetry $U \mapsto A_1 U A_2^\dag$, where $A_1$ and $A_2$ are
constant elements of $SU(2)$. The vacuum $U=1_2$ spontaneously breaks this symmetry
down to the isospin $SU(2)$ subgroup $U \mapsto AUA^\dag$, where $A \in SU(2)$. 
In addition, with the pion mass term present there is a small explicit breaking of
chiral symmetry. The Skyrme model therefore captures the most fundamental property
of QCD, that of spontaneous chiral symmetry breaking.

Using energy and length units of $F_\pi / 4e$ and
$2/eF_\pi$ respectively, the Skyrme Lagrangian can be rewritten as
\begin{equation} \label{eq:l}
L=\int \left\{ -\frac{1}{2}\,\hbox{Tr}\,(R_\mu R^\mu) 
+ \frac{1}{16}\,\hbox{Tr}\,([R_\mu,R_\nu][R^\mu,R^\nu]) 
+ m^2\,\hbox{Tr}\,(U - 1_2) \right\} d^3 x \,,
\end{equation}
where $R_\mu
= (\partial_\mu U)U^{\dag}$, and the dimensionless pion mass
$m = 2m_\pi / eF_\pi$.

At a fixed time, $U$ is a map from $\mathbb R ^3$ to $S^3$, the group manifold of
$SU(2)$. The boundary condition $U \rightarrow 1_2$
implies a one-point compactification of space, so that topologically $U$ can be 
regarded as a map from $S^3$ to $S^3$. As $\pi_3(S^3) = {\mathbb{Z}}$, Skyrme field 
configurations fall into homotopy classes 
labelled by an integer $B$, the baryon number, which is equal to the integral over 
space of the baryon density $B_0(\mathbf{x})$:
\begin{equation}
B = \int B_0(\mathbf{x}) \, d^3 x\,,
\end{equation}
where 
\begin{equation} \label{eq:cur}
B_\mu(\mathbf{x})=\frac{1}{24\pi^2}\,
\epsilon_{\mu\nu\alpha\beta}\,\hbox{Tr}\, \partial^\nu
UU^{\dag}\partial^\alpha U U^{\dag}\partial^\beta U U^{\dag} \,
\end{equation}
is the baryon current.

Restricting to static fields $U({\mathbf{x}})$, the Skyrme energy functional
derived from the Lagrangian is
\begin{equation}
E=\int\left\{-\frac{1}{2}\,\hbox{Tr}\,(R_iR_i)-
\frac{1}{16}\,\hbox{Tr}\,([R_i,R_j][R_i,R_j])-\,
m^2\hbox{Tr}(U-1_2)\right\}d^3x\,.
\end{equation}
For a given baryon number $B$, we denote the minimized energy by ${\cal{M}}_B$, and 
we call the field that minimizes $E$ a Skyrmion. ${\cal{M}}_B$ can be identified with
the static Skyrmion mass. Occasionally, we also refer to some non-global minima of the energy
and some low-lying saddle point solutions
as Skyrmions too. The 
Skyrme energy functional has a nine-dimensional symmetry group, consisting of 
translations and rotations in $\mathbb R ^3$, together with isospin transformations.
Consequently, Skyrmions lie on orbits of this symmetry group. Generically, this orbit is 
nine-dimensional, although for especially symmetric Skyrmions it is of lower dimension.

\section{Symmetric Skyrmions}
The minimal energy Skyrmion in the $B=1$ sector is spherically symmetric and takes 
the form
\begin{equation}
U(\mathbf{x}) = \hbox{exp}\,(if(r)\hat{\mathbf{x}} \cdot
\boldsymbol{\tau})=\cos f(r)1_2 + i\sin f(r)\hat{\mathbf{x}} \cdot
\boldsymbol{\tau}\,,
\end{equation}
where $f$ is a radial profile function obeying an ordinary differential equation
with the boundary conditions
$f(0)=\pi$ and $f(\infty)=0$. Skyrmions with $B > 1$ all
have interesting shapes; they are not spherical like the $B=1$ Skyrmion.
Figure 1 shows surfaces of constant baryon density for Skyrmions 
with $1 \leq B \leq 8$, with the dimensionless pion mass parameter $m=0$ \cite{manton}. 
The surfaces of constant energy density are qualitatively rather 
similar. The $B=2$ Skyrmion, for example, has axial symmetry and its baryon density has a toroidal structure
\cite{b21,b22,b23}. The Skyrmions presented in Fig. 1 have only discrete symmetries
for $B>2$. The $B=3$ and $B=4$ Skyrmions have tetrahedral and octahedral symmetry respectively.
The $B=5$, 6 and 8 Skyrmions have extended dihedral symmetries, and the $B=7$ Skyrmion has icosahedral
symmetry.

Some of these symmetric Skyrmions can be formed instantaneously during the
collision of well separated $B=1$ Skyrmions. For example, three Skyrmions initially 
placed on the vertices of a large contracting equilateral triangle
scatter through the tetrahedral $B=3$ Skyrmion, which then splits into 
a single Skyrmion and a $B=2$ torus \cite{scatter}. The dynamics is 
remarkably similar to the scattering of three $SU(2)$ Bogomolny-Prasad-Sommerfield 
monopoles.

\begin{figure}[h!]
\begin{center}
\includegraphics[width=12cm]{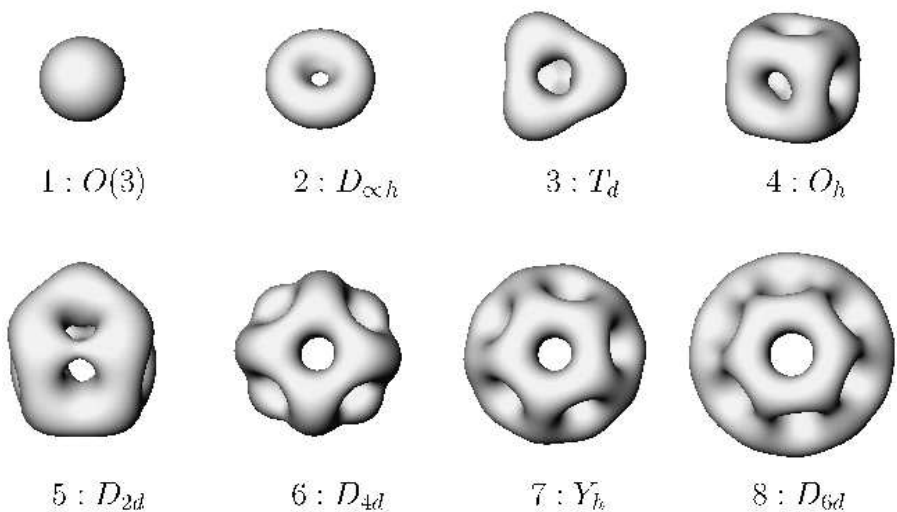}
\caption{Skyrmions for $1\le B\le 8$, with $m=0$. A surface of constant baryon
density is shown, together with the baryon number and symmetry.}
\end{center}
\end{figure}

In order to apply the Skyrme model to nuclear physics, a suitable parameter set must be 
chosen. A key consideration is the value of the dimensionless pion mass parameter $m$.
When $m=0$, the Skyrmions with $B \geq 3$, up to $B=22$ \cite{BS3a,BS3b,BS3c} and
beyond \cite{BHS}, are hollow polyhedra. 
The baryon density is concentrated in a shell of roughly constant
thickness, surrounding a region in which the
energy and baryon density are very small. Such a hollow structure is acceptable for small $B$, but clearly
disagrees with the rather uniform baryon density observed in the interior of
larger real nuclei. However, in the interior region of these Skyrmions the value of $U$ is close to $-1_2$, and for positive
values of $m$, this is the value of $U$ with 
highest potential energy. Not surprisingly, therefore, it is found that
for baryon numbers $B \geq 8$ the hollow polyhedral Skyrmions do not
remain stable when the pion mass parameter $m$ is of order 1 \cite{bs2,bs3}.

New stable Skyrmion
solutions with baryon number a multiple of four have recently been found \cite{bms}.
These solutions are
clusters of cubic $B=4$ Skyrmions, and so
make contact with the $\alpha$-particle model of nuclei \cite{BFWW}. 
They are of more uniform density than the old solutions.
For example, when $m=0$ the $B=8$ Skyrmion is a hollow polyhedron with $D_{6d}$ 
symmetry. However, when $m=1$ the stable solution
is found to be a bound configuration of two $B=4$ cubic Skyrmions, with $D_{4h}$ symmetry
(see Fig. 2). This matches the known physics
that beryllium-8 is an almost bound state of two $\alpha$-particles.
For $B=12$, the new solution is an equilateral triangle of three $B=4$ cubes.
The lowest energy solution has $C_3$ symmetry and is shown in Fig. 3, but there is a solution of very slightly higher
energy with a larger $D_{3h}$ symmetry.
Rearranged solutions are
analogous to the rearrangements of the $\alpha$-particles which model excited
states of nuclei. An example is the Skyrme model analogue of the three
$\alpha$-particles in a chain configuration for an excited state of
carbon-12 \cite{FSW,Mor}, which is displayed in Fig. 4.

\begin{figure}[h!]
\begin{center}
\includegraphics[width=5cm]{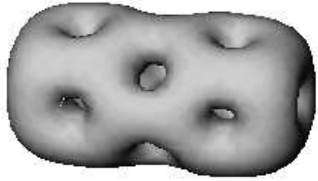}
\caption{Baryon density isosurface for the $B=8$ Skyrmion with $m=1$, resembling two touching
$B=4$ Skyrmions.}
\end{center}
\end{figure}

\begin{figure}[h!]
\begin{center}
\includegraphics[width=9cm]{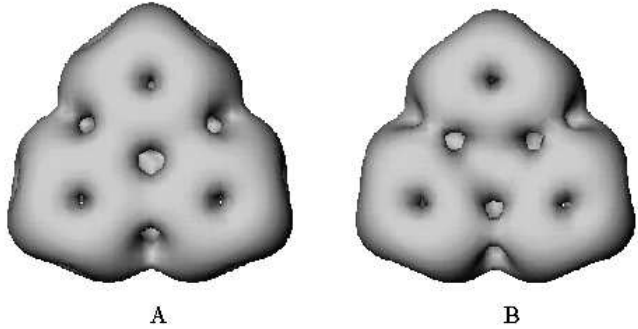}
\caption{Top and bottom views of the $B=12$ Skyrmion with 
triangular symmetry.}
\end{center}
\end{figure}

\begin{figure}[h!]
\begin{center}
\includegraphics[width=6.5cm]{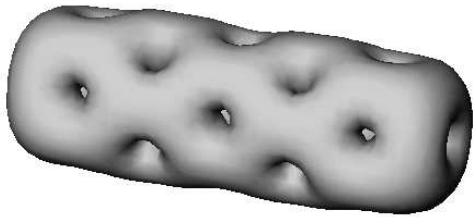}
\caption{$B=12$ Skyrmion formed from three cubes in a line.}
\end{center}
\end{figure}

\section{The Rational Map Ansatz}
Skyrmion solutions are known for several values of $B$, but 
they can only be obtained numerically. Motivated by similarities between 
lumps, monopoles and Skyrmions, an underlying connection in terms of rational maps 
between Riemann spheres was investigated by Houghton, Manton and Sutcliffe \cite{houghton}, and 
this leads to a method of constructing good approximations to several known 
Skyrmions for $m=0$, and also for non-zero $m$. The rational maps have exactly the same symmetries as the Skyrmions
in almost all cases.

A rational map is a map from $S^2$ to $S^2$, which can be expressed 
as
\begin{equation}
R(z)=\frac{p(z)}{q(z)}\,, 
\end{equation}
where $p$ and $q$ are polynomials in $z$. Via stereographic projection,
a point on $S^2$ having complex coordinate $z$ is identified as having
conventional spherical polars given by
\begin{equation}
z = \hbox{tan}\frac{\theta}{2}\,e^{i\phi}\,,
\end{equation}
or equivalently as being the unit Cartesian vector
\begin{equation}
\frac{1}{1+|z|^2}(2\,\text{Re}(z),2\,\text{Im}(z),1-|z|^2)\,.
\end{equation}
We denote a point in $\mathbb R ^3$ by its coordinates $(r,z)$, where $r$ is the
radial distance from the origin. 
The rational map ansatz for the Skyrme field is constructed from a rational map $R(z)$ and
a profile function $f(r)$ as
\begin{equation}
U(r,z) =
\left(
\begin{array}{cc}
\cos f + i\sin f \frac{1-|R|^2}{1+|R|^2} & i\sin f \frac{2\bar{R}}{1+|R|^2} \\
i\sin f \frac{2R}{1+|R|^2} & \cos f - i\sin f \frac{1-|R|^2}{1+|R|^2}
\end{array}
\right)
\,.
\end{equation}
For this to be well-defined at the origin and infinity, one imposes 
$f(0)=\pi$ and $f(\infty)=0$.

The rational map ansatz leads to some simplifications. The baryon number 
is given by
\begin{equation}
B = \int \frac{-f'}{2\pi^2}{\left({\frac{\hbox{sin }f}{r}}\right)}^2\left( \frac{1 +
| z|^2 }{ 1 + |R|^2} \left\vert\frac{dR }{ dz }\right\vert
\right)^2\,\frac{2i\,dz\,d\bar{z}}{(1+|z|^2)^2}\,r^2\,dr \,,
\end{equation}
and this reduces to 
the topological degree of the rational map, which is the greater of 
the algebraic degrees of $p$ and $q$. The energy is given by 
\begin{equation} \label{eq:e}
E=4\pi \int_{0}^{\infty} \left(r^2 f'^2 + 2B\sin^2
f(f'^2+1)+ \mathcal{I}\,\frac{\sin^4 f}{r^2} + 2m^2r^2(1-\cos
f)\right) dr \,,
\end{equation}
in which $\mathcal{I}$ denotes the angular integral
\begin{equation}
\mathcal{I} = \frac{1}{4\pi} \int
\left(\frac{1+|z|^2}{1+|R|^2}\left\vert\frac{dR }{ dz }\right\vert
\right)^4\frac{2i\,dz\,d\bar{z}}{(1+|z|^2)^2}\,.
\end{equation}
Minimal energy solutions within this ansatz are 
found by first minimizing $\cal{I}$ over all maps of degree $B$. 
The profile function $f$ is then found by solving the second order ordinary
differential equation that 
is the Euler-Lagrange equation obtained from (\ref{eq:e}) with $B$, $m$ and $\cal{I}$ 
as parameters. Table 1 lists the energy-minimizing rational maps, their symmetries and the total
Skyrmion energy for 
$1 \leq B \leq 8$ and $m=0$, and in Fig. 5 we present a graph of their corresponding profile functions.
The Skyrmions increase in size with increasing $B$.

\begin{table}[ht]
\centering
\begin{tabular}{c c c c}
\hline \\
$B$ & $R(z)$ & $E/12\pi^2$ & Symmetry \\[0.5ex]
\hline \\
1 & $z$ & 1.23 & $O(3)$ \\[2ex]
2 & $z^2$ & 2.42 & $D_{\infty h}$ \\[2ex]
3 & $\frac{\sqrt{3}iz^2-1}{z(z^2-\sqrt{3}i)}$ & 3.55 & $T_d$ \\[2ex]
4 & $\frac{z^4 + 2\sqrt{3}iz^2 + 1}{z^4 - 2\sqrt{3}iz^2 + 1}$ & 4.55 & $O_h$ \\[2ex]
5 & $\frac{z(z^4+bz^2+a)}{az^4-bz^2+1}$ & 5.74 & $D_{2d}$ \\[2ex]
6 & $\frac{z^4 + ic}{z^2(icz^4 + 1)}$ & 6.82 & $D_{4d}$ \\[2ex]
7 & $\frac{7z^5+1}{z^2(z^5-7)}$ & 7.75 & $Y_h$ \\[2ex]
8 & $\frac{z^6 - id}{z^2(idz^6 - 1)}$ & 8.94 & $D_{6d}$ \\[2ex]
\hline
\end{tabular}
\caption{Energy-minimizing rational maps. The parameters $a$, $b$, $c$ and $d$
are numerically determined as 3.07, 3.94, 0.16 and 0.14 respectively.}
\end{table} 

\begin{figure}[h!]
\begin{center}
\includegraphics[width=10cm]{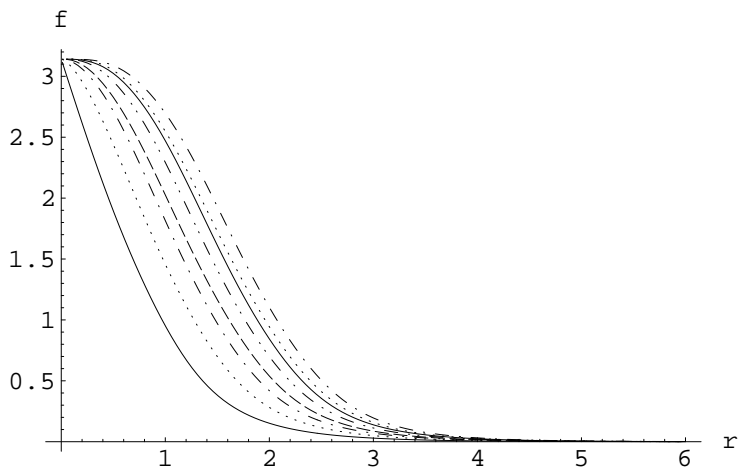}
\caption{The profile functions $f(r)$ for
$B=1$ to 8. $B$ increases from left to right.}
\end{center}
\end{figure}

\section{Skyrmion Quantization}
In the semiclassical method of Skyrmion quantization, one quantizes the spin and 
isospin rotational degrees of freedom while treating the Skyrmion as a
rigid body. As mentioned previously, the symmetry group of the model is 
nine-dimensional. Given a generic static Skyrmion $U_0$, there is a nine-parameter 
set of configurations, all degenerate in energy, obtained from 
$U_0$ by some combination of translation, rotation and isorotation:
\begin{equation}
U(\mathbf{x})=A_1U_0(D(A_2)(\hbox{\bf x}-\mathbf{X}))A_1^{\dag} \,,
\end{equation}
where $A_1, A_2 \in SU(2)$ and $D(A_2)_{ij} = \frac{1}{2}
\hbox{Tr}(\tau_iA_2\tau_j A_2^{\dag}) \in SO(3)$. 
The quantization procedure promotes the collective coordinates 
$A_1, A_2, \mathbf{X}$ to dynamical variables, each depending on time \cite{bc}. 
In what follows, the translational degrees of freedom are ignored and 
the Skyrmions are quantized in their rest frames.

The ansatz for the dynamical Skyrme field is then given by 
\begin{equation} \label{eq:dyn}
\hat{U}(\hbox{\bf x},t)=A_1(t)U_0(D(A_2(t))\hbox{\bf x})A_1(t)^{\dag}\,.
\end{equation}
Inserting this into the
Lagrangian (\ref{eq:l}), we obtain the kinetic energy
\begin{equation} \label{eq:t}
T=\frac{1}{2}a_i U_{ij} a_j - a_i W_{ij} b_j + \frac{1}{2}b_i V_{ij}b_j \,, 
\end{equation}
where $b_i$ and $a_i$ are the angular velocities in space and isospace 
respectively, 
\begin{equation}
a_j = -i\,\hbox{Tr}\,\tau_j A_1^{\dag}\dot{A}_1\,,\,\,\,b_j =
i\,\hbox{Tr}\,\tau_j \dot{A}_2 A_2^{\dag} \,,
\end{equation}
and the inertia tensors $U_{ij}$, $W_{ij}$ and $V_{ij}$ are
functionals of the Skyrmion $U_0$ given by
\begin{eqnarray}
U_{ij} &=& -\int \hbox{Tr}\,
\left(T_iT_j + \frac{1}{4}[R_k,T_i][R_k,T_j]\right)
\, d^3 x \label{eq:u}\,,\\
W_{ij} &=& \int \epsilon_{jlm}\,x_l\,\hbox{Tr}\,
\left(T_iR_m + \frac{1}{4}[R_k,T_i][R_k,R_m]\right) \, d^3 x \label{eq:w}\,,\\
V_{ij} &=& -\int \epsilon_{ilm}\,\epsilon_{jnp}\,x_lx_n\,
\hbox{Tr}\,\left(R_mR_p + \frac{1}{4}[R_k,R_m][R_k,R_p]\right) \, d^3 x \label{eq:v}\,,
\end{eqnarray} 
where $R_k = (\partial_k U_0)U_0^{\dag}$ and 
$T_i = \frac{i}{2}\left[\tau_i,U_0\right]U_0^{\dag}$.

The momenta corresponding to $b_i$ and $a_i$ are the body-fixed spin and 
isospin angular momenta $L_i$ and $K_i$:
\begin{eqnarray}
L_i &=& -W_{ij}^{\rm{T}} a_j + V_{ij}b_j\,,\\
K_i &=& U_{ij} a_j - W_{ij}b_j\,.
\end{eqnarray}
The space-fixed spin and isospin angular momenta are denoted $J_i$ and $I_i$ 
respectively. One regards $L_i$, $K_i$, $J_i$ and $I_i$ as quantum operators, 
each set satisfying the $\mathfrak{su}(2)$ commutation relations. The quantum Hamiltonian
is obtained by re-expressing (\ref{eq:t}) in terms of $L_i$ and $K_i$. The symmetries
of the inertia tensors are related to the symmetries of the Skyrmion $U_0$.
In several cases, these tensors are proportional to the identity matrix, in which case the
Hamiltonian is that of a spherical top. If the matrices have two or three distinct eigenvalues,
the Hamiltonian is that of a symmetric or asymmetric top, respectively. 

Finkelstein and Rubinstein showed that it is possible to quantize a single 
Skyrmion as a fermion by defining wavefunctions on the covering space of the 
classical configuration space, which is a double cover for any value of $B$ \cite{fr}. 
The wavefunction is defined in such a way that it has opposite signs on the 
two points of the covering space that cover one point in the configuration 
space. The basic Finkelstein-Rubinstein (FR) constraints on Skyrmion states for $B>1$ 
implement the requirements of the Pauli 
exclusion principle (for nucleons). In particular they imply that for even $B$ the spin and 
isospin are integral, and for odd $B$ they are half-integral. Further FR constraints 
arise whenever the Skyrmion has special nontrivial symmetries. These constraint equations
are now relatively easy to determine with the help of the rational map ansatz \cite{krusch}.
For example, the toroidal symmetry
of the $B=2$ Skyrmion leads to FR constraints which imply that the quantum ground state
has spin 1 and isospin 0, in agreement with the
quantum numbers of the deuteron \cite{bc}. Also, the FR constraints imply that the lowest quantum state
of the double cube $B=8$ Skyrmion has spin 0 and isospin 0, which is consistent with
the quantum numbers of beryllium-8 \cite{bms}.

A rational map $R(z)$, and hence the 
corresponding Skyrmion, has a rotational symmetry if it satisfies an equation 
of the form $R(M_2(z))=M_1(R(z))$, where $M_1$ and $M_2$ are M\"{o}bius 
transformations.
$M_2$ corresponds to a spatial rotation defined by an angle $\theta_2$ and an 
axis $\mathbf{n}_2$; and $M_1$ corresponds to an isorotation defined by an angle $\theta_1$ 
and an axis $\mathbf{n}_1$. 
Such a symmetry gives rise to a loop in configuration space (one thinks of this as a loop
by letting the isorotation angle increase from 0 to $\theta_1$, while the rotation 
angle increases from 0 to $\theta_2$). The symmetry leads to the following constraint on the wavefunction:
\begin{equation}
e^{i\theta_2\mathbf{n}_2\cdot\mathbf{L}}
e^{i\theta_1\mathbf{n}_1\cdot\mathbf{K}}|\Psi\rangle
= \chi_{\rm{FR}}|\Psi\rangle \,, 
\end{equation}
where the FR sign $\chi_{\rm{FR}}$ enforces the fermionic quantization condition:
\begin{equation} 
\chi_{\rm{FR}} = \left\{ \begin{array}{ll} +1 & \textrm{if
the loop induced by the symmetry is contractible,} \\ -1 &
\textrm{otherwise.} \end{array} \right. 
\end{equation}
Krusch showed that the FR sign depends 
only on the angles $\theta_1$ and $\theta_2$ (provided a crucial base point condition 
is satisfied for further details of which we refer the reader to Ref. \cite{krusch}) 
through the formula
\begin{equation}
\chi_{\rm{FR}} = (-1)^{\cal{N}},\,\,\,\,\, \hbox{where}\,\,\, {\cal{N}}=\frac{B}{2\pi}(B\theta_2 - \theta_1).
\end{equation}

A convenient basis for the wavefunctions is given by the direct products
$|J,J_3,L_3\rangle \otimes |I,I_3,K_3\rangle$, 
which is effectively shorthand for tensor products of Wigner $D$-functions
parametrized by the rotational and isorotational Euler angles. In what follows, 
the arbitrary third components of the space and isospace angular momenta are 
omitted, so basis states are denoted $|J,L_3\rangle \otimes |I,K_3\rangle$.
By considering the reflection symmetries of the rational maps, one can determine the 
parities of the quantum states, but we do not discuss this further here.

\section{Reparametrizing the Model}
Adkins, Nappi and Witten first quantized the $B=1$ Skyrmion and showed that the 
lowest energy states may be identified with the proton/neutron isospin 
doublet \cite{an,anw}. The masses of the nucleons and deltas were used to calibrate the model,
and they obtained these values for the Skyrme parameters:
\begin{equation} \label{eq:param}
e=4.84,\,\,\,F_\pi = 108\,\hbox{MeV}\,\,\,
\hbox{and}\,\,\,m_\pi=138\,\hbox{MeV}\,\,\, 
(\hbox{which implies}~m=0.528)\,. 
\end{equation}
However, the delta is rotating at relativistic speeds and decays very rapidly, so this
calibration is not very reliable \cite{spinning}.

In \cite{mantonwood} we considered the lowest lying quantum state of the $B=6$ Skyrmion, 
which has the quantum numbers of the lithium-6 nucleus in its ground state, 
i.e. spin 1 and isospin 0. We calculated a number of its static 
properties, dependent only on the Skyrme model parameters, and then chose 
$e$, $F_\pi$ and $m$ such that the predictions of the model agree precisely 
with the experimentally determined values. It appears that this new parameter 
choice more accurately describes properties of small nuclei than the traditional parameter set (\ref{eq:param}).
By roughly doubling $m$ to 1.125, we obtained
a mean charge radius of the quantized $B=6$ Skyrmion in close agreement with 
that of the lithium-6 nucleus. This sets a new Skyrme length scale. We then 
showed, provided a slight modification of the rational map is performed (while 
preserving its symmetry), that the quadrupole moment agrees with experiment. 
Finally, by equating the mass of the quantized $B=6$ Skyrmion to the lithium-6 
nucleus, we fitted the energy scale of the model. Keeping the pion mass fixed at 
its physical value, we have a new set of Skyrme parameters:
\begin{equation}
e=3.26,\,\,\,F_\pi = 75.2\,\hbox{MeV}\,\,\, 
\hbox{and}\,\,\,m_\pi=138\,\hbox{MeV}\,\,\, (\hbox{which implies}~m=1.125) \,. 
\end{equation}
Reconsidering the $\alpha$-particle and deuteron as quantized $B=4$ and $B=2$ 
Skyrmions gives further support for these new values. In what follows, 
this new parameter set is used throughout.

\section{Energy Spectra of Light Nuclei}
Here we review and slightly extend our work on
the semiclassical quantization of Skyrmions with $B=4$, 6 and 8, as 
approximated by the rational map ansatz \cite{mmw}. By exploiting the holomorphic character of the 
rational map one obtains useful general expressions for the elements 
of the inertia tensors (\ref{eq:u},\ref{eq:w},\ref{eq:v}) in terms of the approximating rational map. 
These formulae are slightly simpler than those obtained by Kopeliovich \cite{kop}.
Using these formulae, 
one can apply techniques detailed in Ref. \cite{landau} to calculate the energy
spectra of the quantized Skyrmions.

\subsection{$B=4$}
The $B=4$ Skyrmion has octahedral symmetry and a cubic shape, and is described by the 
rational map
\begin{equation}
R(z)=\frac{z^4+2\sqrt{3}iz^2 +1}{z^4-2\sqrt{3}iz^2 +1}\,.
\end{equation}
The generating symmetries of the rational map are
\begin{equation}
R(iz)=\frac{1}{R(z)}\,,\,\,\,\,\,\,R\left(\frac{iz+1}{-iz+1}\right) = e^{i\frac{2\pi}{3}}R(z)\,,
\end{equation}
which lead to the FR constraints
\begin{equation}
e^{i\frac{\pi}{2}L_3} e^{i\pi K_1}|\Psi \rangle = |\Psi
\rangle\,,\,\,\,\,\,
e^{i\frac{2\pi}{3\sqrt{3}}(L_1+L_2+L_3)} e^{i\frac{2\pi}{3}K_3}
| \Psi \rangle = |\Psi \rangle\,.
\end{equation}
Solving these in the basis described 
previously, one obtains the ground state $|0,0\rangle \otimes |0,0\rangle$ with spin 0 and isospin 0, 
which are the quantum 
numbers of the helium-4 nucleus, or $\alpha$-particle. This is in agreement with earlier work of Walhout \cite{walhout}.
The next lowest lying states are a spin 2, isospin 1 state given by 
\begin{equation}
\Big(|2,2\rangle +\sqrt{2}i |2,0\rangle + |2,-2\rangle\Big)\otimes |
1,1\rangle - \Big(|2,2\rangle -\sqrt{2}i |2,0\rangle + |2,-2\rangle\Big)
\otimes |1,-1\rangle\,,
\end{equation}
and a spin 4, isospin 0 state given by
\begin{equation}
\left(|4,4\rangle +\sqrt{\frac{14}{5}}|4,0\rangle + |
4,-4\rangle\right)\otimes |0,0\rangle\,.
\end{equation}

The symmetries imply that the elements of the inertia tensors
$U_{ij}$, $V_{ij}$ and $W_{ij}$ are all diagonal, with $U_{11}=U_{22}$, $V_{ij}=v\delta_{ij}$ and $W_{ij}=0$.
$U_{11}$, $U_{33}$ and $v$ are numerically determined. The quantum Hamiltonian is given by
\begin{equation}
T = \frac{1}{2v}\mathbf{J}^2 + \frac{1}{2U_{11}}\mathbf{I}^2 + \frac{1}{2}\left(\frac{1}{U_{33}}-\frac{1}{U_{11}}\right)K_3^2\,.
\end{equation}
The eigenvalue of $J^2$ in 
states of spin $J$ is $J(J+1)$, the standard result. Similarly $I^2$ has 
eigenvalues $I(I+1)$. The energy eigenvalues are calculated as:
\begin{eqnarray}
E_{J=0,\,I=0} &=& {\cal{M}}_4 = 3679\hbox{\,MeV}\,,\\
E_{J=2,\,I=1} &=& {\cal{M}}_4 + 28.7\hbox{\,MeV} = 3708\hbox{\,MeV}\,,\\
E_{J=4,\,I=0} &=& {\cal{M}}_4 + 39.4\hbox{\,MeV} = 3718\hbox{\,MeV}\,.
\end{eqnarray}
For the ground state, the energy is simply the static mass of the 
Skyrmion, ${\cal{M}}_4$. Comparing this to the mass of the $\alpha$-particle, 3727\,MeV, we see that our 
prediction comes to within 2\% of the experimental value. For the spin 2, isospin 1 state, 
the excitation energy is 28.7\,MeV. We note that hydrogen-4, helium-4 and lithium-4 form an isospin 
triplet, whose lowest energy state has spin 2 and has an average excitation 
energy of 23.7\,MeV relative to the ground state of helium-4, so here the Skyrmion picture 
works well \cite{energy4}. Finally, we predict a spin 4, isospin 0 state with an excitation energy of
39.4\,MeV. Such a state of helium-4 has not yet been seen experimentally. However, 
suggestions that such a state exists with an excitation energy of 24.6\,MeV have been made 
previously \cite{spin4,spin4other}.
In Fig. 6 we present the energy level diagram for the quantized $B=4$ Skyrmion
and the corresponding experimentally observed states.
\begin{figure}[h!]
\begin{center}
\includegraphics[width=7cm]{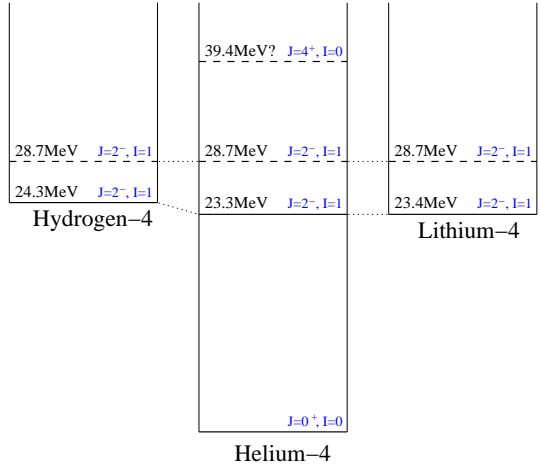}
\caption{Energy level diagram for the quantized
$B=4$ Skyrmion. Solid lines indicate experimentally observed
states, while dashed lines indicate our predictions.}
\end{center}
\end{figure}

\subsection{$B=6$}
The quantization of the $B=6$ Skyrmion was first considered by Irwin \cite{irwin}.
The Skyrmion has $D_{4d}$ symmetry and can be approximated using
the rational map
\begin{equation}
R(z) = \frac{z^4 + ia}{z^2(iaz^4 + 1)}\,,\,\,\,a=0.16\,.
\end{equation}
Solving the FR constraints arising from the Skyrmion's symmetry one obtains its allowed quantum 
states \cite{mantonwood}. The lowest three are $|1,0\rangle
\otimes |0,0\rangle$, $|3,0\rangle \otimes |0,0\rangle$ and
$|0,0\rangle \otimes |1,0\rangle$.
The Hamiltonian is given 
by
\begin{equation}
T=\frac{1}{2V_{11}}\left[\mathbf{J}^2-L_3^2\right] +
\frac{1}{2U_{11}}\left[\mathbf{I}^2-K_3^2\right] +
\frac{1}{2(U_{33}V_{33}-W_{33}^2)}\left[U_{33}L_3^2 + V_{33}K_3^2
+ 2W_{33}L_3K_3\right]\,.
\end{equation}
The static Skyrmion mass, ${\cal{M}}_6$, is set to be 5600\,MeV, just below
the mass of the lithium-6 nucleus, to allow for the spin energy 
which is of order 1\,MeV\footnote{This updates Ref. \cite{mantonwood}
where we treated the spin energy as negligible}. 
The energy eigenvalues corresponding to the lowest
three states are then given by
\begin{eqnarray}
E_{J=1,\,I=0} &=& {\cal{M}}_6 + \frac{1}{V_{11}} = {\cal{M}}_6
+ 1.7\,\hbox{MeV} = 5601\,\hbox{MeV}\,,\\
E_{J=3,\,I=0} &=& {\cal{M}}_6 + \frac{6}{V_{11}} = {\cal{M}}_6
+ 10.3\,\hbox{MeV} = 5610\,\hbox{MeV}\,,\\
E_{J=0,\,I=1} &=& {\cal{M}}_6 + \frac{1}{U_{11}} = {\cal{M}}_6
+ 12.1\,\hbox{MeV} = 5612\,\hbox{MeV}\,.
\end{eqnarray}
These states, together with further allowed states,
are displayed on the energy level diagram in Fig. 7. The
experimental energy level diagram is displayed in Fig. 8 \cite{energy567}. 

\begin{figure}[h!]
\begin{center}
\includegraphics[width=11cm]{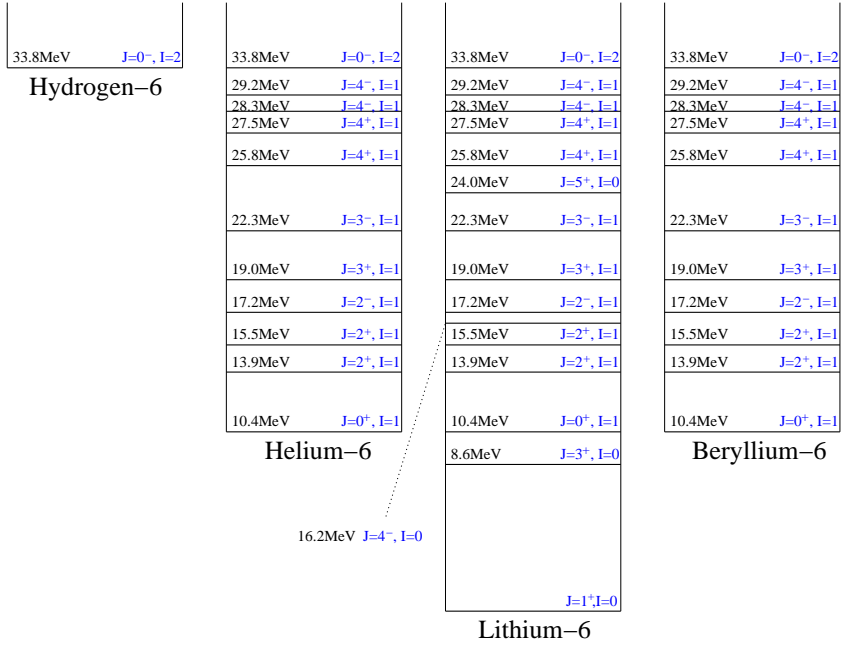}
\caption{Energy level diagram for the quantized $B=6$
Skyrmion. Energies are given relative to the spin 1, isospin 0 ground state.}
\end{center}
\end{figure}

\begin{figure}[h!]
\begin{center}
\includegraphics[width=11cm]{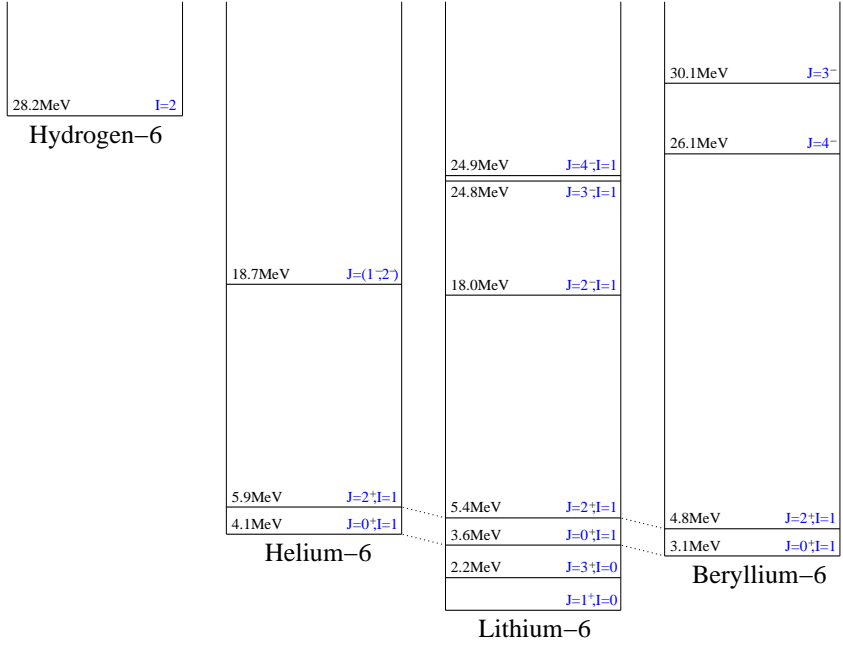}
\caption{Energy level diagram for nuclei with $B=6$.}
\end{center}
\end{figure}

The ground 
state of the lithium-6 nucleus has spin 1 and isospin 0, and there is an 
excited state with spin 3 with excitation energy 2.2\,MeV. 
The lowest states of the isotriplet of helium-6, lithium-6 and beryllium-6
have spin 0 and spin 2. Further states of this isotriplet with 
spins 2, 3 and 4 are experimentally observed although the data are not complete.
One state of an isospin 2 multiplet, the ground state of hydrogen-6, is also
observed. The Skyrme model qualitatively reproduces this spectrum
and has some further states. The hydrogen-6 state is predicted to have spin 0.
The only problem is for the lower spin states 
of lithium-6 as the 
energy splittings are too large by a factor of three to four.
Note that we predict a number of states that have not yet been seen experimentally,
for example spin 4 and spin 5 excited states of lithium-6 with isospin 0. The quantization of further 
degrees of freedom, such as vibrational modes, may lead to improvements of these predictions. For 
example, we have not considered the possibility of the nucleus separating
into an $\alpha$-particle and a deuteron.

\subsection{$B=8$}
As described in section 3, the stable $B=8$ Skyrmion when $m$ is of order 1 resembles two touching $B=4$ cubes
(see Fig. 2). In this case, the rational map ansatz does not provide a quantitatively
accurate approximation. However, a rational map that has the equivalent $D_{4h}$
symmetry can be used to determine the FR constraints and quantum states of the Skyrmion.
Such a rational map is 
\begin{equation}
R(z) = \frac{z^8+bz^6-az^4+bz^2+1}{z^8-bz^6-az^4-bz^2+1}\,,
\end{equation}
with $a$ and $b$ real.
The FR constraints are given by
\begin{equation}
e^{i\frac{\pi}{2}L_3}e^{i\pi K_1}|\Psi\rangle = |\Psi\rangle \,,\,\,\,\,\,
e^{i\pi L_1}|\Psi\rangle = |\Psi\rangle \,.
\end{equation}
These allow a ground state with 
spin 0, isospin 0, an excited state with spin 2, isospin 0, and many more 
excited states. 

To estimate the moments of inertia one can work directly with the known inertia tensors of 
two $B=4$ cubes and as a 
simplifying approximation use the parallel axis theorem \cite{mmw}. The resulting inertia
tensors are diagonal and satisfy
the relations $U_{11}=U_{22}$,
$V_{11}=V_{22}$ and $W_{ij}=0$, and so the Hamiltonian is
\begin{equation}
T =\frac{1}{2V_{11}}\left[\mathbf{J}^2 - L_3^2\right] +
\frac{1}{2U_{11}}\left[\mathbf{I}^2 - K_3^2\right] +
\frac{L_3^2}{2V_{33}} + \frac{K_3^2}{2U_{33}}\,.
\end{equation}

Figures 9 and 10 are energy level diagrams for the quantized $B=8$ 
Skyrmion and for the $B=8$ nuclei \cite{energy8910}, respectively. Our predictions agree well
with experiment. 
The predicted energy of the spin 2, isospin 0 state is 2.9\,MeV, which 
is a very good match to the experimental value of 3\,MeV.
We obtain a spin 2 isotriplet with energy 13.3\,MeV. Experimentally such a 
triplet exists with an average excitation energy of 16.5\,MeV.  We also obtain a spin 3 isotriplet 
with energy 16.2\,MeV, which is experimentally seen with an average excitation energy of
19.0\,MeV. We have also found quintets of $I=2$ states. The lowest of these, with 
spin 0, has been detected experimentally with excitation energies very close 
to our prediction, and includes the helium-8 and carbon-8 ground states. 
We have recently started working 
with the exact values of the inertia tensors and have obtained similar results.

Of particular interest is the prediction from the Skyrme model of a spin 0 isotriplet of
negative parity 
states, which if established experimentally could include new ground states of the lithium-8 and boron-8 nuclei.
Low-lying spin 0, negative parity states could be difficult to observe, as experienced by the difficulty
and the long time taken to observe the bottomonium and charmonium ground state mesons $\eta_b$ and $\eta_c$ \cite{bottom,charm}.

\begin{figure}[h!]
\begin{center}
\includegraphics[width=12cm]{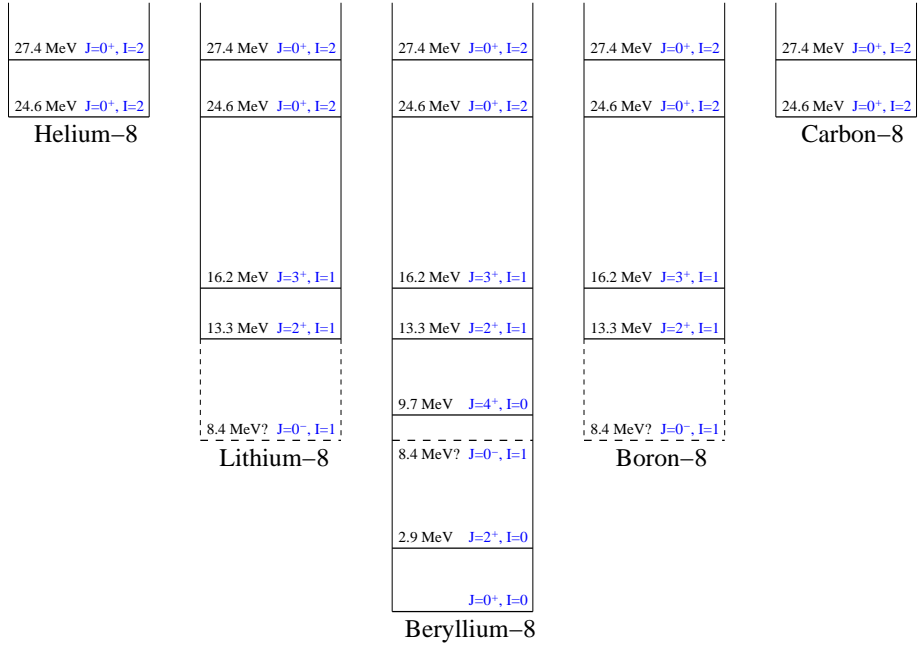}
\caption{Energy level diagram for the quantized $B=8$ Skyrmion. The putative $J=0^-$ isotriplet is represented by dashed lines. 
Higher energy negative parity states are also predicted, but are omitted here.}
\end{center}
\end{figure}

\begin{figure}[h!]
\begin{center}
\includegraphics[width=12cm]{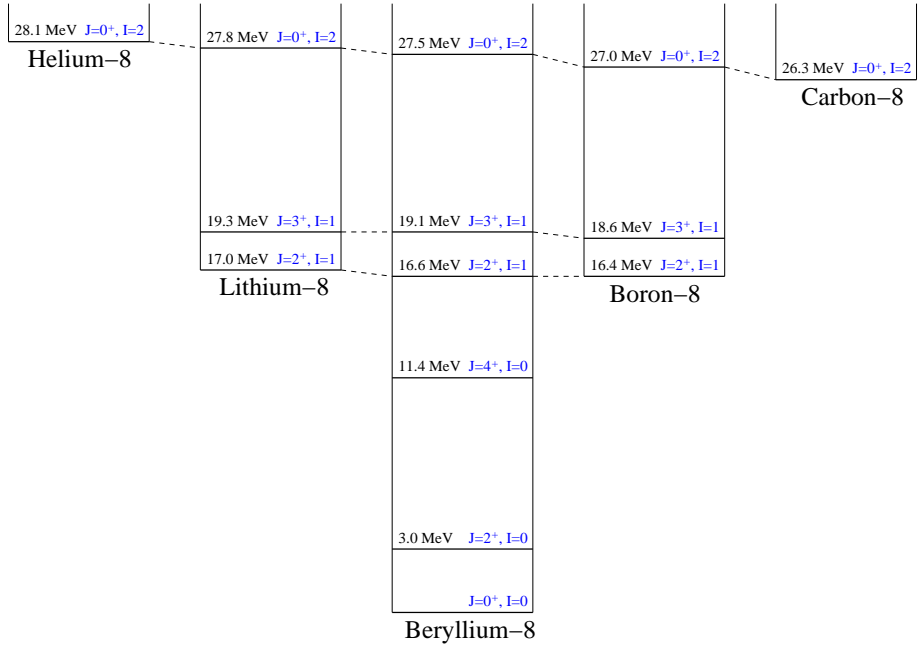}
\caption{Energy level diagram for nuclei with $B=8$ (selected levels).}
\end{center}
\end{figure}

\section{Electromagnetic Form Factors of Quantized Skyrmions}
The results described so far are encouraging for the 
Skyrme model of nuclei. Firstly, stable Skyrmions exist for all 
baryon numbers that we wish to consider. Secondly, they have quantum states 
with the same spin and isospin quantum numbers as their corresponding nuclei. 
And thirdly their energy spectra are in 
reasonable agreement with experiment. A criticism of the model is that nuclei, 
bound states of protons and neutrons, bear little resemblance to highly 
symmetric classical Skyrmions, even taking into account that the classical Skyrmion
occurs in all possible orientations in space and in isospace, with probability density 
determined by the collective coordinate wavefunction.
The comparison of static 
properties and energy levels does not really address this criticism.

Now, the 
internal structure of nuclei can be investigated by electron 
scattering. The finite size of the nuclear charge distribution produces large 
deviations from the differential cross section for scattering from a point 
charge. A measure of this departure is provided by electromagnetic form 
factors. In this section we investigate whether the symmetries of the classical 
Skyrmions give rise to unusual behaviour in the form factors which would be 
incompatible with those of the nuclei they are supposed to model. The form factors of
lithium-6 are calculated here, following the method developed by Braaten and Carson
that was applied to the deuteron \cite{bcform}. We will also consider the charge 
form factor of the $\alpha$-particle.

We need to consider the full moduli space of Skyrmion collective coordinates, 
including translations, so we make the 
ansatz for the dynamical Skyrme field:
\begin{equation}
\hat{U}(\hbox{\bf x},t)=A_1(t)U_0(D(A_2(t))(\hbox{\bf x}-{\hbox{\bf
X}}(t)))A_1(t)^{\dag}\,.
\end{equation}
Form factors are calculated by inserting this
ansatz into the expression for the electromagnetic current
\begin{equation}
{\cal{J}}_\mu =\frac{1}{2}B_\mu + I_\mu ^3\,,
\end{equation}
and determining matrix elements of the resulting operator between quantum states.
$I_\mu ^3$ is the third component of the isospin current density, and $B_\mu$ the
baryon current (\ref{eq:cur}).

The ground state of the quantized $B=6$ 
Skyrmion, which we recall has spin 1 and isospin 0, is
\begin{equation} 
\Psi_{J_3}(\mathbf{p}) = \frac{\sqrt{3}}{8\pi^2}D^1_{0 J_3}(\phi,\theta,\psi)
D^0_{0 0}(\alpha,\beta,\gamma)e^{i\mathbf{p}\cdot \mathbf{X}}
\end{equation}
in terms of Wigner $D$-functions parametrized by the rotational and isorotational 
Euler angles. $J_3$ is the third component of the space-fixed spin, and $\mathbf{p}$
is the momentum.
The charge and quadrupole form factors are defined in the Breit frame 
(i.e. the frame in which the sum of the initial and final momenta is zero) in terms 
of the matrix element of ${\cal{J}}_0$ between ground states:
\begin{equation}
\langle \Psi_{J_3'}(\mathbf{p}')|{\cal{J}}_0(\mathbf{x}=0)|\Psi_{J_3}(\mathbf{p})\rangle
= G_C(q^2) \delta_{J_3'\,J_3} +
\frac{1}{6{\cal{M}}_6^2}G_Q(q^2)\Omega_{J_3'\,a}(3q^aq^b-q^2\delta^{ab})\Omega_{b\,J_3}^\dag\,,
\end{equation}
where $\mathbf{q}=\mathbf{p}'-\mathbf{p}$ is the momentum transfer, $q^2 = \mathbf{q} \cdot \mathbf{q}$, ${\cal{M}}_6$
is the Skyrmion mass and $\Omega_{J_3 a}$ is the unitary matrix relating the Cartesian
basis to the spin 1 angular momentum basis.

Braaten and Carson \cite{bcform}
obtained a general expression for the matrix element of ${\cal{J}}_0$, which 
may be simplified to
\begin{equation*}
\langle \Psi_{J_3'}(\mathbf{p}')|{\cal{J}}_0(0)|\Psi_{J_3}(\mathbf{p})\rangle
= 
\end{equation*}
\begin{equation}
\delta_{J_3'\,J_3} \frac{1}{2} \int j_0(qr)B_0(\mathbf{x}) d^3 x 
+
\Omega_{J_3'\,a}(3q^aq^b-q^2\delta^{ab})\Omega_{b\,J_3}^\dag \frac{1}{4}\frac{1}{q^2}\int (1-3\cos^2 \theta)
j_2(qr)B_0(\mathbf{x})d^3 x\,,
\end{equation}
where $j_n(qr)$ denote spherical Bessel functions 
and $B_0$ is the baryon density of the Skyrmion $U_0$ in its initial orientation.
Therefore
\begin{eqnarray}
G_C(q^2) &=& \frac{1}{2}\int j_0(qr)B_0(\mathbf{x}) d^3 x\,, \\
\frac{1}{{\cal{M}}_6^2}G_Q(q^2) &=& \frac{3}{2}\frac{1}{q^2}\int (1-3\cos^2 \theta)
j_2(qr)B_0(\mathbf{x})d^3 x\,.
\end{eqnarray}
As $q^2 \rightarrow 0$, $G_C(q^2) \rightarrow
3-\frac{1}{2}q^2 \langle r^2 \rangle$ and $G_Q(q^2) \rightarrow {\cal{M}}_6^2Q$,
where $\langle r^2 \rangle$ and 
$Q$ are the squared mean charge radius and quadrupole moment. Note 
that $\langle r^2 \rangle$ is proportional to the derivative of $G_C(q^2)$ with respect to
$q^2$. 

The structure of the lithium-6 nucleus is also described by a magnetic form factor. 
Before going into this, we will firstly describe the calculation of the magnetic 
dipole moment of the quantized $B=6$ Skyrmion in the Skyrme model. The classical magnetic moment is 
defined by
\begin{equation}
\mu_a = \frac{1}{2} \int \epsilon_{abc}x_b {\cal{J}}_c\,d^3 x\,.
\end{equation}
As the ground state 
has isospin 0, the electromagnetic current density ${\cal{J}}_c$ is equal to half the baryon 
current density (\ref{eq:cur}), and so
\begin{equation}
\mu_a  = \frac{1}{4} \int
\epsilon_{abc}x_b B_c\,d^3 x\,.
\end{equation}
Inserting the expression for the rotated classical Skyrmion (\ref{eq:dyn}), we 
obtain 
\begin{equation}
\hat{\mu}_a = D(A_2)^{\rm{T}}_{a\alpha} \left(M_{\alpha k}a_k
+N_{\alpha k}b_k \right)\,,
\end{equation}
where
\begin{eqnarray}
M_{\alpha k} &=& \frac{1}{32\pi^2} \int x_\beta\,\hbox{Tr}\,\left(T_k[R_\alpha,R_\beta]\right) d^3x\,,\\
N_{\alpha k} &=& \frac{1}{32\pi^2} \int \epsilon_{krs} x_\beta x_r\,\hbox{Tr}\,\left(R_s[R_\alpha,R_\beta]\right) d^3x\,,
\end{eqnarray}
and these are calculated for the unrotated Skyrmion.
These matrices are straightforward to calculate approximately using the rational map ansatz.
The $D_{4d}$ 
symmetry of the Skyrmion implies that $M_{\alpha k}$ and $N_{\alpha k}$ are 
diagonal and satisfy $M_{11}=M_{22}=0$ and $N_{11}=N_{22}$.
These relations imply that the terms of $\hat{\mu}_a$ involving $N_{33}$ 
multiply the operator $K_3$, which annihilates the quantum state of the Skyrmion. We ultimately 
obtain
\begin{equation}
\hat{\mu}_a = -\frac{N_{11}}{V_{11}}J_a + \hbox{terms proportional to }K_3\,,
\end{equation}
where $V_{11}$ is a component of the spatial inertia tensor (\ref{eq:v}).
The magnetic dipole moment, $\mu$, is defined to be 
the expectation value of $\hat{\mu}_3$ between quantum ground states with $J_3=1$, 
and so 
\begin{equation}
\mu = -\frac{N_{11}}{V_{11}} = \frac{1}{8V_{11}}\int r^2 (1+\cos^2 \theta) B_0(\mathbf{x}) d^3 x\,.
\end{equation}
$\mu$ is calculated to 
be 0.54\,nm (in physical units), which is quite close to the experimental value for lithium-6 of 0.82\,nm \cite{moment}.

The magnetic form factor is defined in the Breit frame in terms of the matrix 
elements of the spatial components of the electromagnetic current density 
between quantum ground states of the $B=6$ Skyrmion \cite{bcform}:
\begin{equation}
\langle \Psi_{J_3'}(\mathbf{p}')|{\cal{J}}_i(\mathbf{x}=0)|\Psi_{J_3}(\mathbf{p})\rangle
= \frac{1}{2{\cal{M}}_6}G_M(q^2) \Omega_{J_3'\,a}(q^a\delta_i^b-\delta^a_iq^b)\Omega_{b\,J_3}^\dag\,.
\end{equation}
This matrix element is given 
by
\begin{equation}
\langle \Psi_{J_3'}(\mathbf{p}')|{\cal{J}}_i(0)|\Psi_{J_3}(\mathbf{p})\rangle
=\Omega_{J_3'\,a}(q^a\delta_i^b-\delta^a_iq^b)\Omega_{b\,J_3}^\dag
\frac{3}{8V_{11}}\frac{1}{q}\int r\left(1+\cos^2 \theta\right) j_1(qr)B_0(\mathbf{x}) d^3 x\,,
\end{equation}
from which the magnetic form factor is obtained as:
\begin{equation}
\frac{1}{2{\cal{M}}_6}G_M(q^2) = \frac{3}{8V_{11}}\frac{1}{q}\int r\left(1+\cos^2 \theta\right) j_1(qr)B_0(\mathbf{x}) d^3 x\,.
\end{equation}
As $q^2 \rightarrow 0$, $G_M(q^2) \rightarrow 2{\cal{M}}_6\mu$.
So the static electromagnetic properties are recovered from the 
electromagnetic form factors in the limit of 
zero momentum transfer.

In Fig. 11 we present a graph of the absolute normalized charge form factor for lithium-6
and for the quantized $B=6$ Skyrmion.
There is no method of experimentally separating $G_C(q^2)$ and $G_Q(q^2)$.
Due to the smallness of $G_Q(q^2)$, we have made the comparison 
under the assumption that the observed electron scattering cross sections are due entirely to
monopole charge scattering. This is consistent with other approaches \cite{li}.
We observe that the slopes of the theoretical and experimental form factors 
agree at $q^2=0$. This was, of course, to be expected as our new parameter set 
was chosen such that the mean charge radius is correctly predicted. 
The first cusp in the form factor is experimentally seen 
somewhere in the range $7\,{\hbox{fm}}^{-2} \leq q^2 \leq 8\,{\hbox{fm}}^{-2}$. We underpredict the
location of this first 
cusp to be at roughly $2\,{\hbox{fm}}^{-2}$. We recall that the Skyrmion baryon density vanishes at the 
centre of the Skyrmion, which is unlike the conventional picture of a nucleus 
with its baryon density having at most a small dip in the centre. This may be the 
reason why the form factor cusps appear at too low values of the momentum 
transfer. Softening the Skyrmion by allowing it to vibrate may give a better fit.
In comparison, in Fig. 12 we observe that our prediction for the absolute normalized magnetic form factor
agrees rather well with experiment.

\begin{figure}[h!]
\begin{center}
\includegraphics[width=10cm]{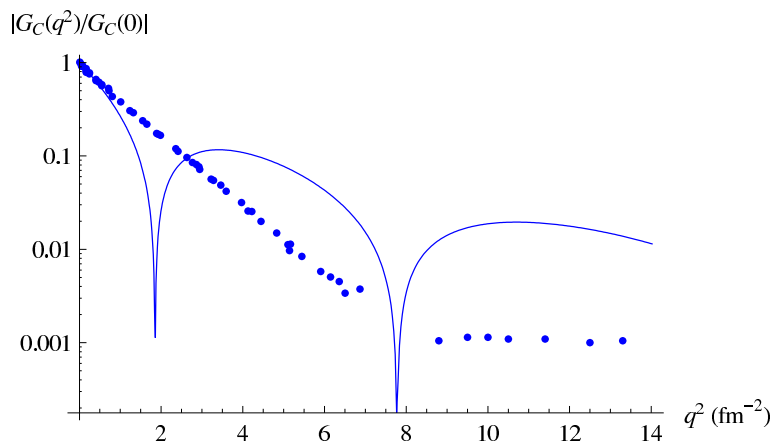}
\caption{Absolute values of the charge form factor of the quantized $B=6$ Skyrmion (solid) compared with
experimental data for lithium-6 (dots) \cite{bumiller,li,suelzle}.}
\end{center}
\end{figure}

\begin{figure}[h!]
\begin{center}
\includegraphics[width=10cm]{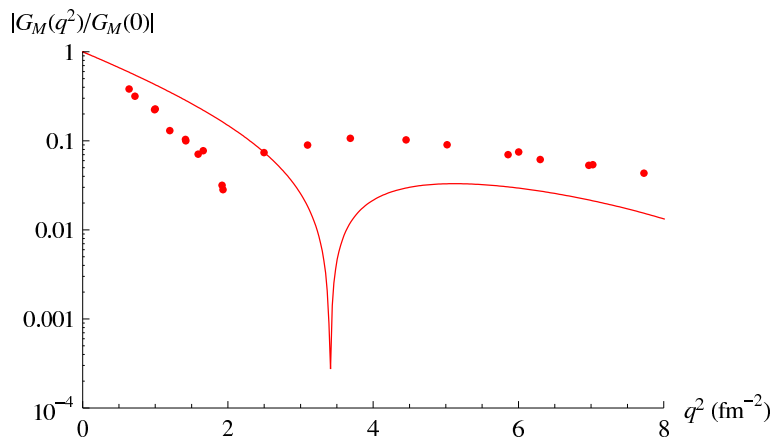}
\caption{Absolute values of the magnetic form factor of the quantized $B=6$ Skyrmion (solid) compared with 
experimental data for lithium-6 (dots) \cite{bergstrom,rand}}
\end{center}
\end{figure}

We turn now to the $\alpha$-particle. 
The cross section for the elastic scattering of electrons off a spin 0 nucleus, 
such as the $\alpha$-particle, depends only on the charge form factor, which 
is given by
\begin{equation}
G_C(q^2) = \frac{1}{2}\int j_0(qr)B_0(\mathbf{x}) d^3 x\,.
\end{equation}
In Fig. 13 we have plotted the absolute normalized values of the charge form factor 
of the $\alpha$-particle and the quantized $B=4$ Skyrmion.
Certainly, 
the two graphs have the same qualitative features, with the appearance of 
cusps in both cases. However, again our predicted cusps appear at smaller values of momentum 
transfer than the experimental cusps. Another thing to note is that we 
overpredict the magnitude of the slope of $G_C(q^2)$ at $q^2=0$, and correspondingly 
overpredict the mean charge radius of the nucleus. A further reparametrization 
of the Skyrme model, simultaneously leading to an accurate mean charge radius and to the 
correct location of the first cusp might be worth investigating. New data on the form factors of light 
nuclei are currently being collected at JLAB and preliminary results indicate 
a second cusp in the charge form factor of the $\alpha$-particle \cite{jlab}.

\begin{figure}[h!]
\begin{center}
\includegraphics[width=10cm]{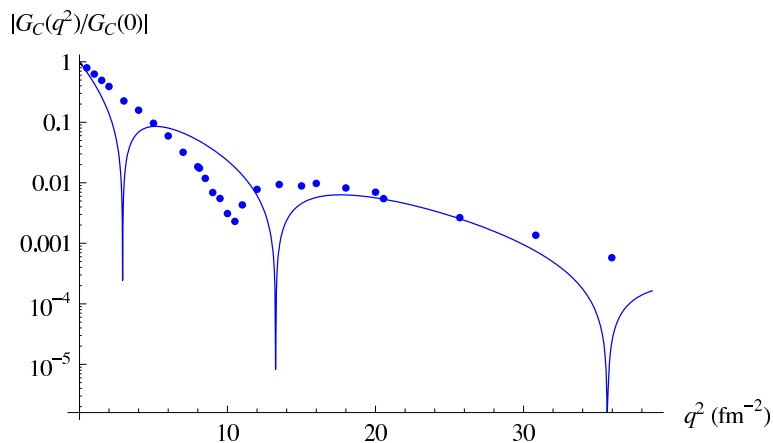}
\caption{Absolute values of the charge form factor for the quantized $B=4$ Skyrmion (solid) 
compared with experimental data for helium-4 (dots) \cite{arnold,frosch}.}
\end{center}
\end{figure}

\section{Conclusion}
The work presented here has certainly provided support for Skyrme's 
field theoretical model for nuclei. In the Skyrme model, the individual $B=1$ Skyrmions merge and
lose their identities in the Skyrmion solutions with $B>1$, 
which is unlike conventional nucleon potential models. This may
capture an essential feature of nuclei in the configurations where nucleons are as close together as possible.
It leads to a new geometry with which to describe
nuclei. Our constraints on the allowed quantum numbers of states provide indirect 
support for this geometrical picture.
The results which we have obtained for energy levels agree well with experiment, and the new 
parameter set which we proposed has enabled us to more accurately estimate 
properties of nuclei over a wider range than before. 
We have made predictions of a number of excited states of nuclei that have not been
seen experimentally. This includes the prediction of a spin 4 state of helium-4, with an
excitation energy higher than those of the experimentally established states. We have also predicted
new spin 0, negative parity ground states of lithium-8 and boron-8.
The qualitative behaviour of our form factors is in quite good agreement with experiment, 
and the symmetries of the classical Skyrmions do not lead to contradictions with 
experiment. 

The approach here is the semiclassical method of quantization, which unifies the treatment of spin and
isospin excitations. We 
have quantized the collective coordinates for translations, rotations and isospin 
rotations, while ignoring further degrees of freedom, which are referred to 
as vibrational modes. Allowing the individual Skyrmions, or subclusters of Skyrmions, to move relative to each other, and
performing a quantization of these degrees of freedom,
would be a significant refinement. We have mainly used the rational map ansatz, which gives 
good approximations for small baryon numbers, but cannot easily be extended 
to higher $B$. With our collaborators Battye and Sutcliffe, we have recently started working with the numerically determined, 
exact Skyrmion solutions including $B=10$ and $B=12$, in order to model the corresponding 
nuclei, including boron-10 and carbon-12. Despite achieving considerable success in describing nuclei with even $B$,
we would like to understand better the odd baryon number sectors of the model.

We have worked with the standard $SU(2)$ 
Skyrme model. The inclusion of strange quarks in the model would be a further
refinement. The introduction of explicit vector meson fields leads to an improved description
of the short-range structure of the nucleons \cite{nyman,zahed}. This could lead to 
further refinement of the modelling of nuclei, but not much
is known about Skyrmions with $B>1$ in these extended models.

Recent work by Sakai and Sugimoto and others has given 
further credence to the idea that at large $N_c$, baryons and nuclei are 
described by some variant of the Skyrme model \cite{ss}. Properties of baryons
were predicted in this framework and it should be possible to predict 
properties of nuclei using this model \cite{hong}.

\end{document}